\documentclass[prl,twocolumn]{revtex4}
\usepackage{tabularx}
\usepackage{amsmath}
\usepackage{amssymb}
\usepackage{graphicx}
\usepackage{dcolumn}
\usepackage{bm}
\usepackage{ulem}
\usepackage{color}

\begin{document}
\title{Universal anyons at the irradiated surface of topological insulator}
\author{Rui Wang$^{1}$}
%\affiliation{National Laboratory of Solid State Microstructures and Department of Physics, Nanjing University, Nanjing 210093, China}
\author{Wei Chen$^{2}$}

%\affiliation{Department of Physics and Centre for Quantum Coherence, The Chinese University of Hong Kong, Shatin, N.T., Hong Kong, China}
\author{Baigeng Wang$^{1}$}
\email{bgwang@nju.edu.cn}
%\affiliation{National Laboratory of Solid State Microstructures and Department of Physics, Nanjing University, Nanjing 210093, China}
\author{D. Y. Xing$^{1}$}
\email{dyxing@nju.edu.cn}

%\affiliation{National Laboratory of Solid State Microstructures and Department of Physics, Nanjing University, Nanjing 210093, China}

\begin{abstract}
Anyons have recently received great attention due to their promising application in topological quantum computation. The best validated system that enjoys the anyonic excitations are the Laughlin states. The quasi-particles in Laughlin states are neither fermions nor bosons but possess the discrete statistical angle $\theta=\pi/m$, with $m$ being an integer. Here we report a possible realization of the universal Abelian anyons, whose statistical angle can be tuned continuously by external parameters and can take any arbitrary values interpolating $\theta=0$ and $\theta=\pi$. The proposed setup is the surface state of a three dimensional topological insulator driven by an amplitude-modulated circularly-polarized light. It is found that the external field leads to a particular Floquet phase, which is a two-spatial-dimensional analogy of the Weyl semimetal phase in the Floquet first Brillouin zone. The chiral anomaly of this phase results in a U(1) Chern-Simons gauge theory with a tunable Floquet Chern number. Owing to this underlying gauge field theory, the irradiated surface of topological insulator constitutes a promising platform for the observation of the universal anyons.
\end{abstract}

\affiliation{$^{1}$ National Laboratory of Solid State Microstructures and Department of Physics, Nanjing University, Nanjing 210093, China \\
$^2$College of Science, Nanjing University of Aeronautics and Astronautics, Nanjing 210016, China\\}
%$^3$Department of Physics, University of California, Davis, One Shields Avenue, Davis, California 95616, USA}

%\pacs{74.45.+c, 74.81.-g, 73.23.-b, 74.78.-w}
% 74.45.+c,   Proximity effects; Andreev reflection; SN and SNS junctions
% 74.81.-g,   Inhomogeneous superconductors and superconducting systems, including electronic inhomogeneities
% 73.23.-b,   Electronic transport in mesoscopic systems
% 74.78.-w,   Superconducting films and low-dimensional structures

\maketitle

The Abelian or non-Abelian Chern-Simons (C-S) gauge field action usually serves as the low energy effective response theory of various topological nontrivial state of matters, such as the chiral spin liquid \cite{Kalmeyer}, the topological insulators (TIs) \cite{Qi} and the Weyl semimetal (WSM) phase \cite{Wan,Zyuzin,Burkov,Chen}.
%For example, the emergent C-S gauge field in the chiral spin liquid leads to the topological order, which manifests itself by the semion excitations and the gapless chiral edge states \cite{Kalmeyer};   The C-S gauge field theory also describes the coupling of electromagnetic field to topological insulators (TIs), and results in the Hall and the topological magneto-electric effect \cite{Qi}; the electromagnetic response of the Weyl semimetal (WSM) phase \cite{Wan} is depicted by a three dimensional generalization of the C-S theory, which predicts the semi-quantized Hall conductance \cite{Burkov} and the chiral magnetic effect \cite{Zyuzin,Chen}.
An interesting application of the C-S gauge field theory is its prediction of the anyonic excitations. As is known, the $1/m$ fractional quantum Hall state (FQH) possesses the anyons with the statistical angle $\theta=\pi/m$. Since $m$ is an integer, the value of $\theta$ is not universal but discrete. Therefore, it is interesting, both theoretically and experimentally, to answer the question: whether anyons with continuously tunable statistics can be realized.
In order to realize the universal tunable anyons, one should search for states that enjoy the Abelian Chern-Simons term with a tunable coefficient \cite{Nayak}. Since the C-S gauge field theory associated with the WSM phase has a tunable coefficient that is proportional to the distance of the Weyl points \cite{Zyuzin}, WSM serves as a promising starting point. However, WSM is defined to be three dimensional material, where the anyons are forbidden \cite{Wen}. These arguments suggest that a two-spatial-dimensional version of the WSM phase, whose electro-magnetic response is described by the $\mathrm{U}(1)$ Abelian C-S theory with a continuously tunable coefficient, constitutes the promising platform to realize the universal anyons.

In order to search for this special phase, we resort to the non-equilibrium states driven by external fields. Recently, the study on the band topology has been extended to the periodic driving case \cite{Sambe}, termed as the Floquet phases. It turns out that the Floquet phases may exhibit nontrivial band topology, and the utilization of the external field also provides more tunable parameters to achieve novel topological phases ~\cite{Kitagawa,Gu,Takuya,Netanel,Eric,Grushin}.  For example, the Floquet fractional Chern insulator\cite{Grushin}, the Floquet topological insulator \cite{Netanel}, the Floquet Weyl semimetal \cite{Rui},  and the Floquet Majorana fermions \cite{Liu,Kundu} are proposed. Moreover, the Floquet-Bloch states of a topological insulator have been observed in experiment\cite{HWang}, which shed light on more future applications. The special feature of the Floquet phases is that the periodically driven system can absorb and emit photons, forming Floquet bands denoted by different photon numbers $n$. This effectively enlarges the ``spacetime" by introducing a fictitious dimension to the undriven phase, and therefore provides a possibility for the two dimensional (2D) driven system to exhibit the chiral anomaly. This fact motivates us to simulate a 3D WSM by driving a certain 2D phase, where the universal anyons may emerge as the quasi-particle excitations.

In this work, we find that the universal anyons can be realized at the irradiated surface of TI. Our main result is that an amplitude-modulated  circularly-polarized light have two effects on the TI surface: (a). Due to the coupling between different Floquet bands, it introduces a fictitious momentum $q$ and enlarges the base manifold of the Hamiltonian, replacing the first Brillouin zone (FBZ) by the Floquet first Brillouin zone (FFBZ) \cite{Platero}.  (b). It generates an effective inhomogeneous Zeeman field, which is a function of $q$. These two effects lead to an undiscovered topological state of matter that exhibits the following non-trivial topological properties. First, it enjoys two gapless Weyl points \cite{Wan} in the FFBZ, which are topologically robust to any perturbations. So we term this state the pseudo Weyl semimetal (PWSM) phase. Second, the PWSM possesses the chiral anomaly, which leads to a U(1) Chern-Simons gauge field theory with a tunable coefficient $C$. This differs from the conventional Chern-Simons action of the integer and fractional quantum Hall state in the sense that it predicts two novel topological behaviors: the universal anyon excitations and the tunable quantum anomalous Hall (QAH) effect.

\textbf{Results}

\textbf{Quasi-energy band structure}

\begin{figure}[tbp]
\includegraphics[width=3.1in]{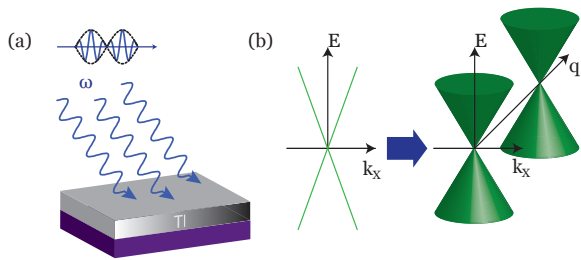}
\caption{(color online) (a) The topological insulator surface driven by an  amplitude-modulated circularly-polarized light. (b) The quasi-energy dispersion (with $k_y=0$) of the driven TI surface. The amplitude-modulated circularly-polarized field drives a single Dirac node into two Weyl nodes that lie in the additional dimension $q$ in the FFBZ, with the location $\pm\mathbf{q}_0$.}
\end{figure}

We motivate the description of the details of the model and its analysis by first showing the main results, which are obtained using a combined formalism of functional path integral and the Floquet-Bloch theory. We consider a TI surface driven by an off-resonant, circularly-polarized light with a slowly-modulated amplitude, as is shown in Fig.1(a). The effect of the off-resonant light, with the period $T_1$, on the TI surface has been studied by Ref.\cite{Kitagawa}, where a constant Zeeman gap is opened, leading to the QAH effect. Here we further require the amplitude $\mathcal{A}(t)$ to be periodic and slowly varying, with the period $T_2$. The function $\mathcal{A}(t)$ is generic, and its specific form does not affect the qualitative results below. For purpose of clarity, we assume $\mathcal{A}^2(q)=|\mathcal{A}_0+\mathcal{A}_1\cos(q)|$, with $q=\omega_2t$ and $\omega_2=2\pi/T_2$.

Using the two-time Floquet formalism (see below), we obtain the effective Floquet Hamiltonian describing the irradiated TI surface \begin{equation}\label{eq12}
  \mathcal{H}_{eff}=\tau^3\sigma\cdot\mathbf{k}+\tau^0\sigma\cdot\mathbf{q}_0.
\end{equation}
Here the $\tau$, $\sigma$ represent the chirality and the band degrees of freedom respectively. In Eq.\eqref{eq12}, we use $\mathbf{k}$ to denote the three dimensional ``momentum", $\mathbf{k}=(k_x,k_y,q)$, where $q$ enters into the Hamiltonian as the introduced ``momentum" due to the external periodic field \cite{Platero}. The quasi-energy dispersion is shown in Fig.1(b). As is shown, the amplitude-modulated circularly-polarized light drives the single Dirac cone in the TI surface into two Weyl nodes (with opposite chiralities) in the low-energy window in the FFBZ at $\pm\mathbf{q}_0$, with $\mathbf{q}_0=(0,0,q_0)$ and $q_0=\pi-\arccos(\mathcal{A}_0/\mathcal{A}_1)$ for $|\mathcal{A}_0/\mathcal{A}_1|\le1$. It is well known that the gapless nodes in the nondegenerate two band model is robust, since all the three Pauli matrices are used up so that no more mass terms can be added to open up the gap. This robustness is due to topology but not symmetry, leading to a peculiar topological matter of state, termed the Weyl semimetal \cite{Wan}. However, different from the conventional 3D WSM, the Weyl nodes here lie in the time axis ($q$) and they are generated by driving the TI surface, therefore we term this peculiar state the PWSM phase. The normal 3D WSM possesses the semi-quantized anomalous Hall effect, where the conductance is proportional to the distance of the Weyl nodes. Here due to the Floquet theory, $k_z$ is replaced by $q$, therefore a similar anomalous Hall conductance proportional to $q_0$ can be expected in the 2D TI surface. Besides, the replacement of $k_z$ by $q$ is nontrivial in the sense that it allows the existence of anyonic excitations. This is because, in this case, no continuous contractible loop that has a vanishing local phase \cite{Wen} can be constructed any more due to the definition of the braiding of the Floquet states \cite{Liu}. To validate these expectations, we perform the calculation of the Berry phase curvature and study the chiral anomaly of the PWSM phase.

\textbf{Berry curvature of the PWSM phase}

For any fixed $k_y$, the Berry phase curvature is defined by the Berry phase gauge field $a_{i}(\mathbf{k})$,
\begin{equation}
f_{xz}(\mathbf{k})=\frac{\partial a_z(\mathbf{k})}{\partial k_x}-\frac{\partial a_x(\mathbf{k})}{\partial k_z},
\end{equation}
with $a_i(\mathbf{k})=-i\sum_{\alpha\in occ}<\alpha\mathbf{k}\mid\partial_{k_i}\mid\alpha\mathbf{k}>$ and $i=x,z$. $k_z$ is used to denote $q$ for brevity and $\mid\alpha\mathbf{k}>$ is the Bloch function of the $\alpha$ band. We calculated and plotted the density distribution of $f_{xz}(\mathbf{k})$ on the FFBZ, as is shown in Fig.2. The first, second and third panel shows $f_{xz}(\mathbf{k})$ for $\mathcal{A}_0/\mathcal{A}_1=-1, 0, 1$, respectively. For $\mathcal{A}_0/\mathcal{A}_1=-1$, we have $q_0=0$, therefore both the two Weyl nodes lie at $(0,0,0)$. When $\mathcal{A}_0/\mathcal{A}_1$ is varied from $-1$ to $1$, the Weyl nodes gradually get separated in the FFBZ, and finally merge with each other at $(0,0,\pi)$. One can check that the $q$ coordinates of the dividing lines between the dark and light areas in Fig.2 coincide with the locations of the Weyl nodes. As such, the Berry phase curvature has nontrivial values inside and trivial values outside the two Weyl points.  This is consistent with the well-known Chern number $C$ of the conventional 3D WSM phase, where $C=1$ and $C=0$ for areas between and outside the two Weyl nodes respectively. The fact that the PWSM shares the same Chern number distribution with the conventional WSM phase is important, since it allows one to modulate the Berry phase accumulation by tuning the amplitude $\mathcal{A}_0/\mathcal{A}_1$. Let us consider the braiding of two quasi-particles. In a complete period from $q=-\pi$ to $q=\pi$, the non-trivial Berry phase accumulation only occurs in the topological non-trivial (light) region, while the contribution is zero for the trivial (dark) region. Therefore, the total Berry phase in a complete period is proportional to the width of the light area, that is proportional to the distance of the Weyl points, $\theta\propto2q_0\propto q_0$. In this sense, the amplitude brings about the tunable Berry phase via modulating the separation between Weyl points. This understanding shows that the the time-evolution of the Floquet states can result in the universal anyon statistics (see below), and the anyon here emerges due to completely different reasons from that of the FQH effect. In the following, we strictly prove the existence of the universal anyons by studying the underlying gauge field theory.
\begin{figure}[tbp]
\includegraphics[width=3.4in]{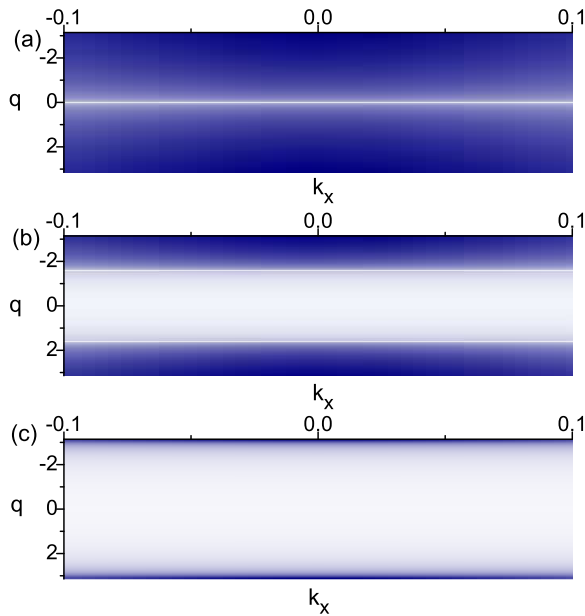}
\caption{(color online) The density distribution of the Berry phase curvature $f_{xz}(\mathbf{k})$ in the FFBZ (with $k_y$ fixed to 0.1). The first, second, third panel depicts the case where $\mathcal{A}_0/\mathcal{A}_1=-1,0,1$ respectively.}
\end{figure}

\textbf{Pseudo chiral anomaly}
To extract the universal anyons, we study the electromagnetic response theory of the PWSM phase. The external electromagnetic field $A_{\mu}$ is minimally coupled to the pseudo Weyl fermions, \textit{i.e.}, Eq.\eqref{eq12}. Utilizing the chiral symmetry, we can eliminate the distance of the Weyl fermions and arrive at a massless Dirac fermion \cite{Zyuzin}. However, when performing the chiral transformation, great attention must be paid to the Jacobian of the integral measure, which can lead to the chiral anomalies and some observable effects \cite{Fujikawa}.

The problem of the chiral anomaly of the conventional WSM is well studied \cite{Fujikawa,Zyuzin}. In the proposed PWSM phase, since all the Dirac matrices are well-defined, a similar chiral anomaly should also exist. With a few modifications, we arrive at the action $\delta S$ describing the chiral anomaly (see Method),
\begin{equation}\label{eq16}
  \delta S=\frac{C}{4\pi}\int d^2rdt\epsilon^{\mu\nu\rho}A_{\mu}\partial_{\nu}A_{\rho},
\end{equation}
where $\mu,\nu,\rho=0,1,2$ and $e$, $\hbar$ is set to 1.  Eq.\eqref{eq16} is a $\mathrm{U}(1)$ C-S theory with a tunable coefficient $C$, with $C=q_0/\pi$. To make the physical meaning more explicit, it is convenient to make a Hubbard-Stratonovich decomposition, which leads to
\begin{equation}\label{eq3n}
  \mathcal{L}=-m\frac{1}{4\pi}a_{\mu}\partial_{\nu}a_{\rho}\epsilon^{\mu\nu\rho}+\frac{1}{2\pi}A_{\mu}\partial_{\nu}a_{\rho}\epsilon^{\mu\nu\rho},
\end{equation}
where $m=1/C$ and $a_{\mu}$ is the introduced auxiliary field. The above equation exactly coincides with the effective Lagrangian of the $\nu=1/m$ FQH \cite{Nayak} and it is interesting to observe that the distance of Weyl nodes is the analogy of the filling factor in the FQH state. The key difference here is that $\mathcal{A}_0/\mathcal{A}_1$ is not discrete but can be tuned continuously. A variation of the action  $\delta S$  shows that this state enjoys a tunable QAH effect \cite{Zyuzin}, with the conductance
$\sigma_{xy}=\frac{Ce^2}{h}$. This is very similar to the semi-quantized Hall conductance in the WSM phase, but the Hall current here resides in the 2D TI surface and can be adjusted by $\mathcal{A}_0/\mathcal{A}_1$. The Hall conductance is shown by the brown curve in Fig.3, which varies between $0$ and $e^2/h$.

Eq.\eqref{eq16} indicates a field theory of anyons. This well-known conclusion is discussed in Ref.\cite{Semenoff,Frohich,Polyakov}. Despite the detailed discussion, the following arguments can show the anyons clearly. From the equation of motion, $\frac{\delta\mathfrak{L}}{\delta A_0}=0$, we know that flux are attached to point particles, forming quasi-particle excitations \cite{Nayak}. The statistical angle for unit $a_{\mu}$ charge is extracted to be the Berry phase $\theta$ accumulated during the braiding of the quasi-particles \cite{Wen}. In this way, we obtain that $\theta=C\pi=q_0$, in agreement with the conclusion in the last section.  As is discussed, $C$ can be continuously tuned by the amplitude $\mathcal{A}_0/\mathcal{A}_1$, so we arrive at the universal anyonic excitations with tunable $\theta$ at the irradiated TI surface, where the anyons are bound states of charge and flux. The interesting universal tunable statistics is shown by the blue curve in Fig.3. As one varies $\mathcal{A}_0/\mathcal{A}_1$ from $-1$ to $0$ and then to $1$, the quasi-particles can evolve from bosons to semions and then to fermions.
%Besides, anyons with arbitrary statistical angel can be prepared singly by adjusting the external field amplitude. This suggests possible applications in the topological quantum computation.
\begin{figure}[tbp]
\includegraphics[width=3.4in]{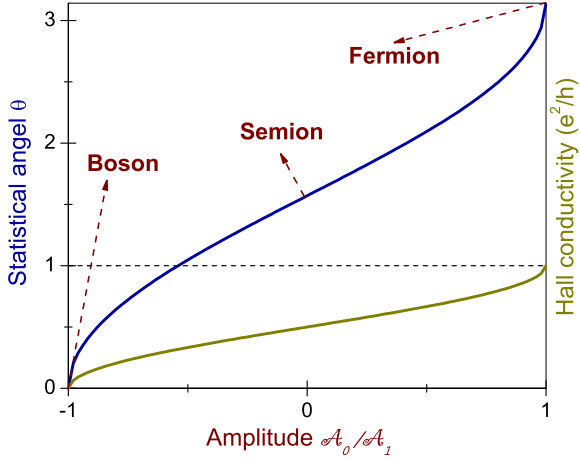}
\caption{(color online) The calculated statistical angle $\theta$ and the Hall conductance versus the amplitude $\mathcal{A}_0/\mathcal{A}_1$. }
\end{figure}

\textbf{The model Hamiltonian}

The Hamiltonian describing the driven TI surface reads
\begin{equation}\label{eqq11}
  \mathcal{H}_{\mathbf{k}}(\mathcal{A}(t),t)=\sigma\cdot\mathbf{k}_{\parallel}+(\sigma_x\mathcal{A}(t)\cos\omega_1t+\sigma_y\mathcal{A}(t)\sin(\omega_1t+\pi/2)),
\end{equation}
where $v_F$ is set to $1$ for brevity and $\mathbf{k}_{\parallel}=(k_x,k_y)$ is the two-dimensional lattice momentum. Since the Dirac cone is a low-energy effective description, an energy cutoff $\Lambda$ is implicit. As required, we have $\mathcal{A}(t+T_2)=\mathcal{A}(t)$, $\mathcal{H}_{\mathbf{k}}(\mathcal{A}(t),t+T_1)=\mathcal{H}_{\mathbf{k}}(\mathcal{A}(t),t)$.
In the following, we focus on the case where $\omega_1=2\pi/T_1$ lies in the high frequency regime with $\omega_1\gg\Lambda$, (see Ref.\cite{Kitagawa}), and $\omega_2$ lies in the low frequency regime, with $\omega_2\ll\Lambda$, as required by Ref.\cite{Platero}. Due to the presence of the two time scale $T_1$, $T_2$ ($T_1\ll T_2$), it is appropriate to resort to the two-time formalism of the Floquet theory \cite{Breuer}, which leads to
\begin{equation}\label{eqq3}
 \widetilde{\mathcal{H}}_{\mathbf{k}}(A(t),\widetilde{t})\Phi_{\mathbf{k}}(t,\widetilde{t})=i\partial_{t}\Phi_{\mathbf{k}}(t,\widetilde{t})
\end{equation}
where $\widetilde{\mathcal{H}}_{\mathbf{k}}(\mathcal{A}(t),\widetilde{t})=\mathcal{H}_{\mathbf{k}}(\mathcal{A}(t),\widetilde{t})-i\partial_{\widetilde{t}}$. $\widetilde{t}$ indicates the fast time and $t$ denotes the slow time. In the two-time formalism, $\widetilde{t}$ can be viewed as a parameter \cite{Liu}. Moreover, since $\omega_2$ lies in the low frequency regime, we can use the Floquet-Bloch theorem and then drop out the term ``$i\partial_{t}$" \cite{Platero}, so that for fixed $\widetilde{t}$, the Hamiltonian is defined in the FFBZ. The Floquet state $u_{\mathbf{k},t}(\widetilde{t})$ and the quasi-energy are determined by
$\widetilde{\mathcal{H}}_{\mathbf{k},t}(\widetilde{t})u_{\mathbf{k},t}(\widetilde{t})=E u_{\mathbf{k},t}(\widetilde{t})$. In the frequency $\omega_1$ space, the Floquet operator $\widetilde{\mathcal{H}}_{\mathbf{k},t}(\widetilde{t})$ is mapped into,
$\widetilde{\mathcal{H}}_{\mathbf{k},t;m,n}=\mathcal{H}_{\mathbf{k},t;mn}-n\omega_1\delta_{m,n},
$
with
\begin{equation}\label{eqq12}
 \mathcal{H}_{\mathbf{k},t;mn}=\frac{1}{T_1}\int^{T_1}_0 d\widetilde{t} \mathcal{H}_{\mathbf{k},t}(\widetilde{t})e^{i\omega_1(m-n)\widetilde{t}}-n\omega_1\delta_{m,n}.
 \end{equation}
Since $\omega_1$ lies in the off-resonant high frequency regime, we cannot drop out the term ``$n\omega_1\delta_{n,m}$". Instead, a perturbation treatment can be performed  \cite{Grushin,Takuya,Kitagawa}, which leads to, $\mathcal{H}_{\mathbf{k},t}\simeq \mathcal{H}_{\mathbf{k},t;0,0}+\frac{1}{\omega_1}[ \mathcal{H}_{\mathbf{k},t;0,1}, \mathcal{H}_{\mathbf{k},t;0,-1}]$. Inserting the Hamiltonian of the TI, Eq.\eqref{eqq11}, we obtain the effective Floquet Hamiltonian describing the irradiated TI surface in the second quantized form,
\begin{equation}\label{eqq13}
  H=\frac{1}{T_2}\int^{T_2}_0 dt d\mathbf{k}_{\parallel}c^{\dagger}_{\mathbf{k}_{\parallel},t}(\sigma\cdot \mathbf{k}_{\parallel}+\frac{1}{\omega_1}\mathcal{A}^2(t)\sigma_z)c_{\mathbf{k}_{\parallel},t},
\end{equation}
The Hamiltonian shows that the amplitude-modulated circularly-polarized light has two effects. (a). It enlarges the base manifold of the Hamiltonian, replacing the FBZ by the FFBZ. (b) It generates an effective tunable Zeeman field, which opens and closes the gap of the TI surface. Further assuming a specific function $\mathcal{A}(t)$, the Hamiltonian can be reduced to the pseudo Weyl fermions in the low energy window, \textit{i.e.}, Eq.\eqref{eq12}. 

Now we are going to discuss the experimental applicability of our results. First, $\omega_1\gg\Lambda$ and $\omega_2\ll\Lambda$ should be satisfied. The topological surface states of the realistic material such as $\mathrm{Bi}_2\mathrm{Se}_3$, $\mathrm{Bi}_2\mathrm{Te}_3$ and $\mathrm{Sb}_2\mathrm{Te}_3$ have been well investigated \cite{Hzhang}. Take $\mathrm{Sb}_2\mathrm{Te}_3$ as example, the cutoff energy of the surface Dirac cone is estimated to be $\Lambda\approx0.1eV$ \cite{Hzhang}. So, the frequencies $\nu_1=\omega_1/(2\pi)$ and $\nu_2=\omega_2/(2\pi)$ should satisfy $\nu_1\gg25\mathrm{THz}$ and $\nu_2\ll25\mathrm{THz}$.   Second, we stress the requirement on the amplitude $\mathcal{A}(q)$. Taking into account $e$, $\hbar$ and $v_F$, the perturbation treatment in Eq.\eqref{eqq13} is correct only when $\mathcal{A}_m^2e^2v^2_F/\omega_1\hbar\ll \Lambda$, where $\mathcal{A}_m$ is the maximum value of $\mathcal{A}(q)$. %Assuming $\mathcal{A}^2(q)=\mathcal{A}_0+\mathcal{A}_1\cos(q)$ as before, we have $A_m=(\mathcal{A}_0+\mathcal{A}_1)^{1/2}$.
We introduce a dimensionless number $\overline{\mathcal{A}}_m=e\mathcal{A}_ma/h$, with $a$ being the lattice constant of $\mathrm{Sb}_2\mathrm{Te}_3$. Then one can estimate that $\overline{\mathcal{A}}_m$ should satisfy $\overline{\mathcal{A}}_m\ll0.06$ (for $\nu_1=250\mathrm{THz}$). So, we do not have particular requirement on the amplitude strength as long as it is not too large. Last, we would like to remark that the splitting of the pseudo Weyl nodes $q_0$ depend neither on the absolute value of the amplitude nor on the frequencies. It only depends on the ratio of $\mathcal{A}_0$ and $\mathcal{A}_1$, which are parameters that can be tuned experimentally. From $\mathcal{A}_0/\mathcal{A}_1=-1$ to $\mathcal{A}_0/\mathcal{A}_1=1$, the splitting of Weyl nodes can be easily tuned from 0 to $\pi$. So, we conclude that the Weyl node separation can be achieved and modulated as long as the condition on the frequencies $\nu_1$ and $\nu_2$ are satisfied.

\textbf{Discussion}
Formally, the coupling constant of the topological action (such as $C$ in Eq.\eqref{eq16}) can be identified as a topological invariant in terms of the Berry fiber bundle \cite{Volovik}. Now we consider the topological invariant and try to reveal the physical essence of the PWSM phase. In order to do so, we recall the effective gauge field theory describing the electromagnetic response of the Chern insulator \cite{Qi}, which is very similar to Eq.\eqref{eq16}. The only difference is that $C$ is replaced by the first Chern number $C_1$. Since $C_1$ is calculated by the Berry phase defined in terms of the Bloch state  \cite{Qi} and the Bloch state is the counterpart of the Bloch-Floquet state in the PWSM phase, we generalize the first Chern number to its Floquet version using the two-time Floquet theory \cite{Liu,Breuer},
\begin{equation}\label{eq4n}
  C_F=\frac{1}{2\pi}\oint d\mathbf{R}\ll u(\mathbf{R},t)\mid i\partial_{\mathbf{R}}\mid u(\mathbf{R},t)\gg,
\end{equation}
where $\mathbf{R}$ is a varying parameter, $\mid u(\mathbf{R},t)>$ is the Bloch-Floquet state and $\ll\cdot\mid\cdot\gg=(1/T_2)\int^{T_2}_0dt<\cdot\mid\cdot>$ is the generalized inner product \cite{Sambe}. If one views the PWSM phase at any fixed parameter $t$, the PWSM phase would be either a 2D normal insulator or a Chern insulator with the first Chern number $C_1=0$ and $C_1=1$ respectively, depending on whether $t$ lies in the trivial or non-trivial regime (see Fig.2). The evolution of the Floquet state in a complete periodic $T_2$ undergoes two topological phase transitions with the energy gap closes and then reopens. Taking this into account, Eq.\eqref{eq4n} can be calculated to be $C_F=q_0/\pi$, which exactly equals to the coefficient $C$ in Eq.\eqref{eq16}. The identification of $C$ with the Floquet Chern number $C_F$ shows that the obtained action Eq.\eqref{eq16} is actually the effective topological field theory describing a Floquet version of the quantum Hall (QH) state.

%The relation $C=C_F$ also indicates an equivalence between the Berry phase accumulated in the braiding process and the Berry phase defined in terms of the Bloch-Floquet state, since both are extracted to be $\theta=C\pi=C_F\pi$. This equivalence further allows a physical justification of the emergence of the universal anyons: As shown in Fig.3, the Berry phase $\theta=C_F\pi$ defined by Eq.\eqref{eq4n} is proportional to the distance of the Weyl points, which can be continuously tuned by the field amplitude $\mathcal{A}_0$, leading to the tunable statistical angle of the quasi-particle excitations.

Now we can conclude that the TI surface driven by an amplitude-modulated circularly-polarized light is a realization of the Floquet QH state, which shows a Weyl semimetal-like dispersion in the low energy window in the FFBZ. This PWSM phase is a close analogy of the normal 3D WSM, with the $k_z$ lattice momentum replaced by the time dimension $q$. A result of this substitution is that no continuous contractible loop that has a vanishing local phase can be constructed any more using the time dimension, therefore the anyons are in principle permitted. This is validated by the associated gauge field theory, since we find that the chiral anomaly of the PWSM phase brings about a 2+1D $\mathrm{U}(1)$ Chern-Simons theory with a tunable coefficient, which further generates the universal anyonic statistics and the tunable QAH effect.  This finding, to our best knowledge, serves as the first theoretical proposal to support the universal anyons in realistic solid state materials.

\textbf{Method}

\textbf{Combined formalism of functional integral and the Bloch-Floquet theory}

In order to study the chiral anomaly of the PWSM phase, we develop a field theory of the periodically driven state. Since we are interested in the electromagnetic response of the PWSM phase, only the $\mathrm{U}(1)$ gauge field case needs to be considered. We start from the well-proved observation in Ref.\cite{Platero}: A Hamiltonian defined in $n-1$D first  Brillouin zone (FBZ) is equivalent to a static one defined in $n$D FFBZ, if it is driven by a field in the low frequency regime (with the period $T$). The effective static Hamiltonian generally reads, $H_F=\frac{1}{T}\int^T_0 dt\int d\mathbf{k}c^{\dagger}_{\mathbf{k},t}\mathcal{H}(\mathbf{k},t)c_{\mathbf{k},t}$.
This equation shows that $t$ can no longer be treated simply as time, it is a ``momentum" that enlarges the FBZ to FFBZ. To obtain the electromagnetic response of $H_F$, we take three following steps. (a). By straightforward generalizations of the conventional coherent-state path integral representation, we develop an effective action describing this non-equilibrium state, which reads
\begin{equation}\label{eq2n}
  S=\frac{1}{T}\int^{+\infty}_{-\infty} dt\int^T_0 dt^{\prime} \int d\mathbf{k} c^{\dagger}_{\mathbf{k},t,t^{\prime}}[i\partial_t-\mathcal{H}(\mathbf{k},t^{\prime})]c_{\mathbf{k},t,t^{\prime}},
\end{equation}
In deriving Eq.\eqref{eq2n}, a new time coordinate $t^{\prime}$ is introduced due to the driven field. This treatment is essentially a field theory representation of the Floquet Green's function method in Ref.\cite{Martinez,Kitagawa}, and it also agrees with the two-time formalism introduced in Ref.\cite{Breuer}. (b). We introduce a $\mathrm{U}(1)$ gauge field $A_{\mu}$ ($\mu=0,1,...,n$) that couples to the Floquet phase, leading to the action
\begin{equation}\label{eq5}
  S=\int d\tau d\mathbf{r}c^{\dagger}_{\mathbf{r},\tau}[\partial_{\tau}+ieA_0+\mathcal{H}(-i\mathbf{\nabla}+e\mathbf{A})]c_{\mathbf{r},\tau},
\end{equation}
where we have made a wick rotation to the Euclidean spacetime and set the volume of ``real" space to be 1. The driven field leads to a mixed space of lattice coordinate and photon number $\mathbf{r}=(r_1,...,r_{n-1},r_n)$ \cite{Platero} by introducing an extra dimension (denoted by $r_n$).  $\mathbf{A}=(A_1,A_2,...,A_n)$, with $A_n$ being an auxiliary component. $A_{\mu}$ is independent on $r_n$ since the perturbation field $A_{\mu}$ does not rely on the photon number index of the Floquet state.  $-i\mathbf{\nabla}$ is the momentum operator of $k=(k_1,...,k_{n-1},t^{\prime})$. (c). Eq.\eqref{eq5} can be simplified utilizing the symmetry of $S$, followed by integrating out the  matter fields. Transformation of $S$ may bring about nontrivial Jacobian associated to the integral measure of the partition function $Z=\int Dc^{\dagger}Dce^{-S}$, denoted by $\delta S$.  Therefore, the one fermion loop effective action for the gauge field can be obtained
\begin{equation}\label{eq7}
  S_{eff}[A_0,\mathbf{A}]=\mathrm{Tr}[\mathrm{log}(\partial_{\tau}+ieA_0+\widetilde{\mathcal{H}}(-i\mathbf{\nabla}+e\mathbf{A}))]+\delta S,
\end{equation}
where $\widetilde{\mathcal{H}}(-i\mathbf{\nabla}+e \mathbf{A})$ denotes the transformed Hamiltonian.
A perturbative treatment of $A_{\mu}$ can further expand $S_{eff}$, making possible the calculation of the Feynman diagrams order by order. Formally, using both the functional path integral and the Bloch-Floquet theory, we arrive at Eq.\eqref{eq7}, which is the effective gauge field theory describing the electromagnetic response of the Floquet phases under our consideration.

\textbf{Derivation of the pseudo Weyl chiral anomaly}

Using the formalism introduced in first section in the Method, the action of the PWSM phase can written as
\begin{equation}\label{eeq1}
  S=\int d\tau d\mathbf{r}c^{\dagger}_{\mathbf{r},\tau}[\partial_{\tau}+ieA_0+\tau^3\sigma\cdot(-i\nabla+e\mathbf{A})+\tau^0\sigma^3q_0]c_{\mathbf{r},\tau},
\end{equation}
where we have coupled a $\mathrm{U}(1)$ gauge field $A_{\mu}$ to the Floquet phase. Here, as discussed in the last section, $A_3$ is an auxiliary component and $A_{\mu}$ is independent on $r_3$. This action has the chiral symmetry, \textit{i.e.}, $S$ remains unchanged under the transformation $c_{\mathbf{r},\tau}\rightarrow e^{-i\tau^3\theta/2}c_{\mathbf{r},\tau}$. One can show that the action can be simplified to
\begin{equation}\label{eeq2}
  S=\int d\tau d\mathbf{r}c^{\dagger}_{\mathrm{r},\tau}[\partial_{\tau}+ieA_0+\tau^3\sigma\cdot(-i\nabla+e\mathbf{A})]c_{\mathbf{r},\tau},
\end{equation}
if $\theta$ satisfies $\theta(\mathbf{r})=2q_0r_3$. Therefore, the chiral transformation shows that two separate Weyl points can be equivalently shifted into one Dirac node. This observation is physically incorrect. In fact, Eq.\eqref{eeq2} only contributes to the first term in Eq.\eqref{eq7}, and we have missed the anomalous term coming from the integral measure $\delta S$. The particular transport behavior of the pseudo Weyl semimetal state comes from $\delta S$, therefore in the following we give a detailed derivation of $\delta S$, using the Fujikawa's method \cite{Fujikawa,Zyuzin}.

By introducing the Dirac matrices with $\gamma^0=\tau^1$, $\gamma^i=i\tau^2\sigma^i$ ($i=1,2,3$), and $\gamma^5=-i\gamma^0\gamma^1\gamma^2\gamma^3$, one can further write the action into,
\begin{equation}\label{eeq3}
  S=\int d\tau d\mathbf{r}\overline{c}_{\mathbf{r},\tau}[i\gamma^{\mu}(\partial_{\mu}+ieA_{\mu})+i\gamma^3(i\gamma^5q_0)]c_{\mathbf{r},\tau}.
\end{equation}
To calculate the Jacobian of the chiral transformation associated with the functional integral measure. We introduce the infinitesimal chiral transformation $c_{\mathbf{r},\tau}\rightarrow e^{-i\tau^3\theta ds/2}c_{\mathbf{r},\tau}$, which can be denoted as the operator $\hat{U}=e^{-i\tau^3\theta ds/2}$. From the action Eq.\eqref{eeq3}, we can introduce a Dirac kernel
\begin{equation}\label{eeq4}
  \Xi=\gamma^{\mu}[\partial_{\mu}+ieA_{\mu}+ib_{\mu}(1-s)\gamma^5],
\end{equation}
where $b_{\mu}=(0,0,0,q_0)$. Then, under the infinitesimal transformation, the Jacobian of the functional integral measure $\mathcal{J}=\mathrm{Det}(\hat{U}^{-2})\equiv e^{\delta S(s)}$ can be calculated, which leads to
\begin{equation}\label{eeq5}
  \delta S(s)=i\frac{ds}{2\pi}\int d\tau d\mathbf{r}\theta(\mathbf{r})\sum_n\phi^{\star}_n\gamma^5\phi_n,
\end{equation}
where $\phi_n$ is the eigenfunction of the Dirac kernel with the eigenvalue $\epsilon_n$. Since $\gamma^5$ satisfy $\{\gamma^5,\Xi\}=0$, $\gamma^5\phi$ is also a eigenfunction which corresponds to the eigenvalue $-\epsilon_n$. Due to the orthogonal condition, we have $\sum_n\phi^{\star}\gamma^5\phi_n=0$ except for $\epsilon_n=0$. This suggests one to pick out the zero energy eigenstates. This can be achieved by introducing a regularization factor \cite{Fujikawa},
 \begin{equation}\label{eeq6}
\delta S(s)=ids\int d\tau d\mathbf{r}\theta(\mathbf{r})\sum_n\phi^{\star}_n\gamma^5e^{-\Xi^2/M^2}\phi_n,
\end{equation}
with the regularization factor $M$ taking the limit $M\rightarrow\infty$. To calculate the sum of $n$ in Eq.\eqref{eeq6}, we make transformation to the momentum space (FFBZ). Then the total anomalous action can be calculated by further integrating $s$ from $s=0$ to $s=1$, which leads to
\begin{equation}\label{eeq7}
  \delta S=i\int^1_0ds\int d\tau d\mathbf{r}\theta(\mathbf{r})\int \frac{d\omega}{2\pi}\frac{d\mathbf{k}_{\parallel}}{2\pi^2}\frac{dq}{2\pi}tr(\gamma^5e^{-\Xi^2/M^2}).
\end{equation}
Finally, after expanding the exponential, we arrive at
\begin{equation}\label{eeq8}
  \delta S=\frac{1}{32\pi^2}\int dtd\mathbf{r}_{\parallel}dr_3\theta(\mathbf{r})\epsilon^{\mu\nu\rho\sigma}F_{\mu\nu}F_{\rho\sigma},
\end{equation}
where $F_{\mu\nu}=\partial_{\mu}A_{\nu}-\partial_{\nu}A_{\mu}$. A Wick rotation has been performed to return back to the real time space. Since $A_{\mu}$ is independent on $r_3$, we can integrate $r_3$ in the action. Further integrating by parts and then taking into account $\theta(\mathbf{r})=2q_0r_3$, we obtain the action describing the chiral anomaly of the pseudo Weyl semimetal phase. 
\begin{equation}\label{eeq9}
  \delta S=\frac{q_0}{4\pi^2}\int d^2rdt\epsilon^{\mu\nu\rho}A_{\mu}\partial_{\nu}A_{\rho},
\end{equation}
where $\mu$, $\nu$, $\rho=0,1,2$.
%Even though the PWSM phase shares the similar chiral anomaly with the conventional 3D WSM phase \cite{Fujikawa}, one needs several following modifications in the specific calculation. (a). Considering the mixing spacetime of the underlying problem, we utilize a mixing representation of coordinate and ``momentum" , $(r_0,r_1,r_2,q)$. (b). In this basis, the covariant derivative $D_{\mu}$ is redefined to be $D^{\prime}_{\mu}=(D_i,D_3)$, with $D_i=\partial_i+ieA_i$ ($i=0,1,2$) and $D_3=i(q+eA_3)$. The energy-momentum tensor $F_{\mu\nu}$ also modifies accordingly. The axion field becomes an axion operator, $\hat{\theta}=\theta(r_0,r_1,r_2,-i\partial_{q})$. (c) The gauge field theory is reduced to 2+1D due to the independence of $A_{\mu}$ on $r_n$.

\textbf{Acknowledgement}

We wish to acknowledge X. G. Wan, S. Y. Savrasov, L. B. Shao, Hongyan Lu, L. Sheng, Y. M. Pan and H. Q. Wang for valuable discussions. This work was supported by 973 Program under Grant No. 2011CB922103, and by
NSFC (Grants No. 60825402, No. 11023002 and No. 91021003).

\textbf{Author contributions} All authors designed and performed the research and wrote the manuscript.

\textbf{Competing financial interests}

The authors declare no competing financial interests.

\end{document}